\documentclass[a4paper]{article}
\usepackage{graphics}
\begin{document}
    
    \title{
Probing large distance higher dimensional gravity from lensing data.}

\author{S. R. Choudhury$^{a}$\thanks{email: src@ducos.ernet.in}\ , \and
G. C. Joshi$^{b}$\thanks{email: joshi@tauon.ph.unimelb.edu.au}\ , \and
S. Mahajan$^{a}$\thanks{email: sm@ducos.ernet.in}\ , \and and Bruce H. J.
McKellar$^{b}$\thanks{email:
b.mckellar@physics.unimelb.edu.au}\\ \relax \\
$^{a}$ Department of Physics, University of
Delhi,\\ Delhi 110007, India. \\
$^{b}$ School of Physics, University of Melbourne,\\
Victoria, Australia, 3010.
}

\date{15 April 2002}

\maketitle

\vspace{1cm}

\abstract{The modifications induced in the standard weak-lensing
formula if Newtonian gravity differs from inverse square law at large
distances are studied.  The possibility of putting bounds on the mass
of gravitons from lensing data is explored.  A bound on graviton mass,
esitmated to be about 100 Mpc$^{-1}$  is obtained from analysis of
some recent data on gravitational lensing.}

\vspace{1cm}

Ever since the successful unification of weak and e.m. interactions into
an `electroweak' theory, hopes have been raised for extensions of these
ideas to strong interactions ( Grand Unified Theories) and even further to
a full unified theory including Gravitation as well.  One of the foremost
difficulties in incorporating gravity with the electroweak theory
concerns the so called `hierarchy' problem namely the huge difference in
the scales of the electroweak theory which is $\sim$ 1 TeV, with that of
quantum gravity which is much higher at $ 10^{19}$ GeV.  During the last
four years, an attractive idea has been introduced \cite{ark}
to overcome the hierarchy problem based on a higher dimensional space time
scenario.  In the earliest of such approaches, space-time is
(4+n) dimensional with the n-extra spatial dimensions compact.  Matter
through Standard Model (SM) - fields is proposed to be confined to the
4-dimensional slice while gravity being the metric field is all over.  There
is only one basic Planck scale in the theory comparable to the TeV-scale
electroweak scale.  The weakness of gravitational interactions in our
4-dimensional world comes about through the celebrated relation:
\begin{eqnarray}
M_{Pl,4-\mbox{{\scriptsize dim}}}^{2}=M_{Pl,(4+n)-\mbox{{\scriptsize 
dim}}}^{2+n} R^n
\end{eqnarray}
where R is the size of the compact dimensions.  The law of gravity remains
practically unchanged with a 4-dimensional Planck scale for distances
$r>>R$ whereas for $r<<R$ the gravitational law changes to a
$\frac{1}{r^{1+n}}$ potential with a scale determined by $M_{Pl}^{4+n}$.
Thus, as far as the validity of Newton's law is concerned, deviations can
be explored in the small distance region, lower than the current
experimental limits.\cite{hoy}\ \\

Subseqently, Randall and Sundrum(RS) \cite{rs} proposed a somewhat
different
higher dimensional scenario which once again solves the hierarchy problem.
The RS construction is in a total of 5-dimensions, with the fifth
dimension in the form of a compact torus with opposite points identified.
At the fixed points of the $S^1/Z_2$ orbifold, one has two 3-branes
wherein the SM fields are confined at one end and  a 'hidden' world is
confined in the other. It is further assumed that there is a negative
tension in our brane and a positive tension in the other and also a bulk
cosmological constant. With the tensions in the brane and bulk finely
tuned , one can solve the 5- dimensional Einstein equations to get a non-
factorizable metric with an exponential `warp' factor. The five
dimensional Planck scale in this model is comparable to our Planck scale
but the `warp' factor effectively reduces all mass scales in our theory
including the vev of the scalar field in the electroweak theory to the TeV
range. The hierarchy problem is thus solved indirectly. Once again in this
theory there are no deviations of the gravitation law at large distance.\
\\ 

The RS scenario above is in a matter free universe. In the presence of
matter, complications arise in the form of the theory not being able to
reproduce standard cosmological results like the dependence of the Hubble
constant on the matter density. This arises essentially because, in this
model,  we live in the negative tension brane.  It is in this
context, that extensions of the RS scenario have been proposed \cite{kog}
wherein there are more branes and we live in a positive tension brane.  The
most notable feature of these extensions is the possibility of deviations
from Newtonian gravity at large distances. Thus , in the model  of Kogan
et.al., in addition to a massless graviton one also has a massive graviton
with a tiny mass which is coupled strongly relative to the massless one. The
effective Planck scale that we see, $M_{Pl}$, gets related to the Planck
scale of the theory, $M$, by the relation:
\begin{eqnarray}
\frac{1}{M_{Pl}^2}=\frac{1}{M^2}. \big[1 + \frac{1}{w^2}\big]
\end{eqnarray}
where $w$ is the 'warp' factor typical of the RS kind of scenario.  Since 
$w$ is a
small number compared to unity, gravity is essentially dominated by the
graviton with a tiny mass rather than by the massless one. This would
imply that at large distances gravity will fall like an Yukawa potential
rather than a pure  1/r  law. The theory is unable to give any estimate of
the mass of this graviton. Laboratory experiments to put bounds on this of
course are useless since we know that Newtonian gravity works very well at
least upto planetary scales. One thus has to look into possible
cosmological measurables to detect possible violations of Newtonian
gravity law.\ \\ 

In a number of other models which have a modification of gravity at 
large distances \cite{dvali+}, the transition can be modelled as a 
Yukawa modification of the Newtonian potential, and can be analysed as 
if one has a finite graviton mass.

It is in the context of the motivation outlined above that we address the
question of the sensitivity of gravitational lensing measurements to possible
deviations from Newtonian gravity. Admittedly, cosmological theories do
not have the same theoretical precision of theories like the SM but it
will still be useful to know the nature of deviations from Newtonian
gravity that current precision of gravitational lensing data can
accomodate. \ \\ 

In this note we focus our attention to a very recent measurement of
gravitational lensing parameters by a cluster of stars at around an
average redshift of $z=1.2$ \cite{van} to evaluate the compatibility of the
data with a `massive' graviton. We follow the standard treatment of the
relationship between the `lensing' parameters and density fluctuations
\cite{bar}. In the standard gravitational scenario, the power spectrum
$P_{\kappa}(l)$ of the effective convergence is given by, assuming a flat
universe:\cite{bar}
\begin{eqnarray}
P_{\kappa}(l)&=& \frac{9 H_0^4\Omega_m^2}{4c^4} \int_0^{w_H} dw
\frac{W^2(w)}{a^2(w)} P_{\delta}(\frac{l}{w},w)
\end{eqnarray}
In the last equation, $\Omega_m$ is the matter density scaled to the
critical density, $w_H$ is the horizon distance, $a(w)$ is the scale factor
related to the redshift by the relation $a^{-1}=(1+z)$, $W(w)$ is 
related to the 
normalized source distribution function G(w) by
\begin{eqnarray}
W(w)=\int_w^{w_H} dw^\prime G(w^\prime) \big(1- \frac{w}{w^\prime}\big)
\end{eqnarray}
and $P_{\delta}(k,w)$ is the density contrast function at a distance
$w$. For a single source at  $w=w_s$ , we have $$G(w)=\delta(w-w_s)$$ and
$P_{\kappa}$ reduces to
\begin{eqnarray}
P_{\kappa}=\frac{9 H_0^4 \Omega_m^2}{4c^4} \int_0^{w_s} dw
 \big(1 - \frac{w}{w_s}\big)^2 \frac{1}{a^2(w)}
P_{\delta}(\frac{l}{w},w)
\end{eqnarray}
The modification of this last equation, if the graviton has a mass m is
easily obtained by observing that the density contrast function enters the
rhs of the equation through the relation between the gravitational
potential $\Phi$ and the density fluctation $\delta$:
\begin{eqnarray}
\nabla^2 \Phi = \frac{3 H_0^2 \Omega_m^2}{2 a} \delta
\end{eqnarray}
which becomes a simple multiplicative relation in Fourier space. This last
equation gets modified, if the gravitation field  has the Yukawa form
$\frac{e^{-\mu r}}{r}$  instead of a pure 1/r form, into
\begin{eqnarray}
\big(\nabla^2 - m^2\big) \Phi = \frac{3 H_0^2 \Omega_m^2}{2 a} \delta
\end{eqnarray}
so that the power spectrum $P_{\kappa}(l)$ gets modified t
$P_{\kappa}^m(l)$ given by
\begin{eqnarray}
P_{\kappa}^m (l) = \frac{9 H_0^4 \Omega^2}{4 c^4} \int_0^{w_s} dw
\frac{P_{\delta}(\frac{l}{w},w)}{a^2(w)}
\Big[\frac{\frac{l^2}{w^2}}{\frac{l^2}{w^2}+m^2}\Big]^2
\end{eqnarray}
The equations for $P_{\kappa}$ and $P_{\kappa}^m$ are the two basic
results that we shall use to compute the effect of massive gravitons. For
this purpose, we focus our attention on a recent extensive measurement of
lensing data from a cluster at around an average redshift of z=1.2. In
particular we shall concentrate on the measured values of the variance of
the power spectrum $\gamma^2$ smoothed over a filter of radius $\theta$
which is related to the $P_{\kappa}$ by:
\begin{eqnarray}
\gamma^2(\theta) = \frac{2}{\pi \theta^2} \int_0^{\infty} dl
P_{\kappa}(l)  J_1^2(l \theta),
\end{eqnarray}
where for simplicity we ignore the variation of the redshifts of the
sources and assume that the entire cluster is at z=1.2.  For the density
contrast spectrum $P_{\delta}$, we assume a form:
\begin{eqnarray}
P_{\delta}(k,w)&=& N. k  T_k^2(q) a^2(w) \nonumber \\
q&=& \frac{k}{\Omega h^2},
\end{eqnarray}
where N is the normalization constant and $T_k(q)$ is given by \cite{pea}
\begin{eqnarray}
T_k(q)= \frac{log(1+2.34q)}{2.34q}
[1+3.89q+(16.1q)^2+(5.46q)^3+(6.71q)^4]^{-\frac{1}{4}}
\end{eqnarray}
N is determined as usual by a $\sigma_8$ normalization procedure so that
it is determined with the value of $\sigma_8$ as a parameter.
To account for the non-linear evolution, we do not use the expression for
$P_{\delta}(k,w)$ as above directly but use it to determine the
expression
for the density power spectrum evolved according to the HKLM
procedure.\cite{pea} This involves defining an integration variable
$k_{nl}$ in the integrals above and a non-linear power-spectrum
$\Delta^2_{nl}(k_{nl})$ as follows:\cite{pea}
\begin{eqnarray}
\Delta^2_{l}(k_l,w)&=&\frac{k_l^3}{2 \pi^2} P_{\delta}(k,w) \nonumber \\
\Delta^2_{nl}(k_{nl},w)&=& f_{nl} (\Delta^2_l(k_l,w)) \nonumber \\
f_{nl}(x)&=&x \big[ \frac{1+N(x)}{1+D(x)}\big]^\frac{1}{\beta} \nonumber
\\
N(x)&=& B \beta x + \big(A x\big)^{\alpha\beta} \nonumber \\
D(x)&=& \Big[ \frac{(A x)^{\alpha} g^3(\Omega)}{V\sqrt(x)}\Big]^{\beta}.
\end{eqnarray}
The constants A, B, V, $\alpha$, $\beta$ and g are defined as
follows:\cite{dod}
\begin{eqnarray}
A&=& 0.482 (1+ n/3)^{-0.947} \nonumber \\
B&=& 0.226 (1+ n/3)^{-1.778} \nonumber \\
\alpha &=& 3.310 (1+ n/3)^{-0.344} \nonumber \\
\beta&=& 0.862 (1 +n/3)^{-0.287} \nonumber \\
V &=& 11.55 (1 + n/3)^{-0.423} \nonumber \\
g&=&\frac{5}{2} \Omega_m
\big[\Omega_m^{\frac{4}{7}}-\Omega_v+(1+\frac{1}{2}\Omega_m)
(1+\frac{\Omega_v}{70})\big]^{-1}.
\end{eqnarray}
$\Omega$ in the above equations is the sum of the matter density
$\Omega_m$ and the vacuum contribution $\Omega_v$. Finally the spectral
index n used above is determined as a function of k via the equation
\begin{eqnarray}
n = \frac{d log(P_{\delta(k)})}{d log(k)}.
\end{eqnarray}
The numerical calculation  above thus involves four parameters,
$\sigma_8$, h, and $\Omega_m$  assuming a flat universe and also assuming
the form of the Dark matter spectrum given above. At first it seems
hopeless to obtain any useful information for the mass m but it is not
actually so. Looking at the graph of the measured variable $\gamma^2$
versus the angle $\theta$ as shown in Fig.1, it is clear that at low
angles the value of $\gamma^2$ is not affected by the presence of the mass
or otherwise. In other words, we can  use the value of $\gamma^2$ in
effect to fix our parameters. With the parameters fixed thus, we can
estimate the values of the same quantity $\gamma^2$ at larger angles which
now is sensitive to the presence or otherwise of a graviton mass.\ \\

\begin{figure}[ht]
      
    \centerline{\includegraphics{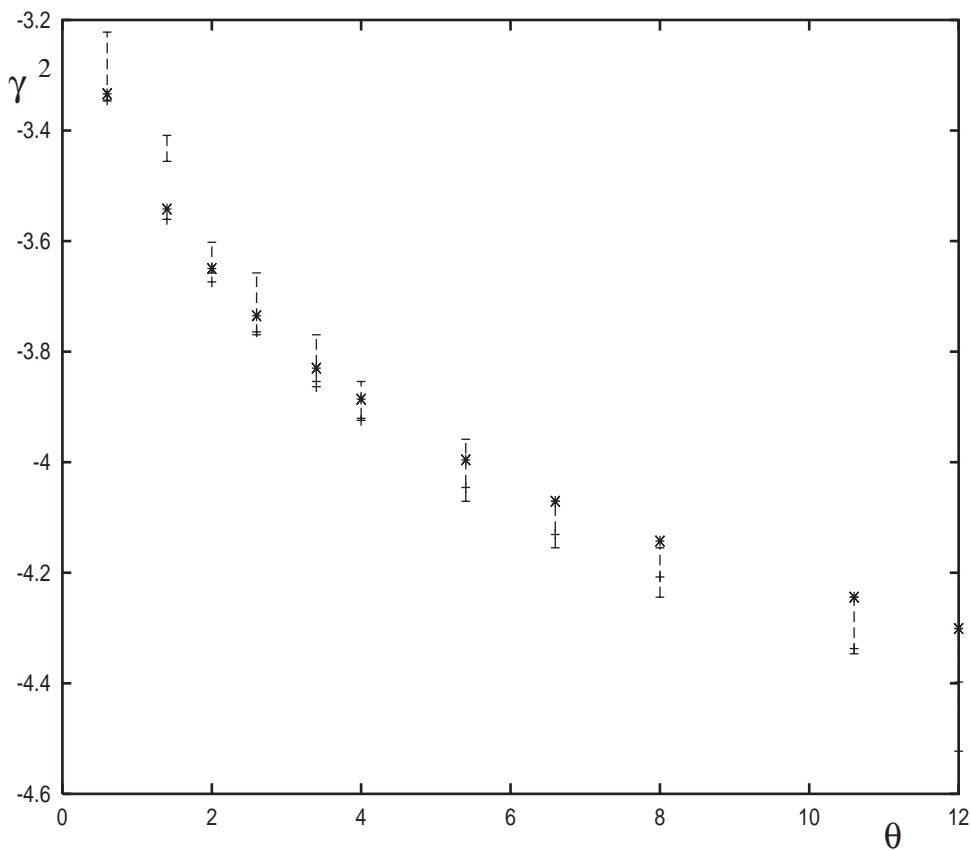}}
    
     \caption{Our fit to the variance data referred to in the text for
   values of h=0.21, $\sigma_8$=0.85 and $\Omega_m$=
   0.3. The  points marked * refer to the usual Newtonian gravity
   whereas the  points marked + uses a modified Newtonian
    law with a mass m chosen to be $m^{-1}$=100 Mpc;
    this value is chosen so that the spread of the points at
    the higher angle values along the x-axis approximately 
    equals the error bars.}\label{fig}
 
\end{figure}

Figure~\ref{fig} summarizes our result for the best fit obtained with values of
h=0.21, $\sigma_8$=0.85 and $\Omega_m=0.3$ in the context of our flat
universe model. Using these, we see that at higher values of the angle,
the precision of the experimental values is compatible with mass m such
that
\begin{eqnarray}
m^{-1} = 100 Mpc
\end{eqnarray}
or higher i.e. masses heavier than about the inverse of 100 Mpc seem to be
ruled out.  A similar limit has been found in an analysis of planetry 
orbits by Gruzinov \cite{gruzinov}.
It has been estimated by Binetruy and Silk \cite{bin} that
current precision of Cosmic microwave background  gives some kind of a
limit once again on the effective graviton mass, which according to their
estimate is
\begin{eqnarray}
m^{-1} \geq r_{h,ls}
\end{eqnarray}
where $r_{h,ls}$ is the horizon distance at last scattering, which is 
of order 3 Mpc for a Hubble Constant of 75 Km s$^{-1}$ Mpc$^{-1}$. Our
estimate, although less direct,  provides a better limit to this
quantity.\ \\ 

In conclusion, gravitational lensing data provides a window for the
detection of deviations from Newtonian gravity at large distances . A
present crude estimate is deviations if at all can occur at distances
beyond 100 Mpc. With more refined data and some more definitive estimates
of various cosmological parameters, this can certainly be improved.

\section*{Acknowledgements}
This work was supported in part by the Australian Research Council. 
SRC would like to thank Professor McKellar and Dr.  Joshi and the
School of Physics for hospitality, where this work was begun.

\end{document}